\documentclass[pre]{revtex4}

\usepackage{amssymb,amsfonts,amsmath}
\usepackage[pdftex]{graphicx}
\usepackage{multirow}
\usepackage{color}
\usepackage{booktabs} 

\begin{document}








\title{Multiscale mobility networks and the large scale spreading of infectious diseases}


\author{Duygu Balcan${}^{1,2}$}
\author{Vittoria Colizza${}^3$}
\author{Bruno Gon\c calves${}^{1,2}$}
\author{Hao Hu${}^{2,4}$}
\author{Jos\' e J. Ramasco${}^{3}$}
\author{ Alessandro Vespignani${}^{1,2}$}

\affiliation{${}^{1}$Center for Complex Networks and Systems Research, School of Informatics and Computing, Indiana University, Bloomington, IN 47408, USA}
\affiliation{${}^{2}$Pervasive Technology Institute, Indiana University, Bloomington, IN 47404 USA}
\affiliation{${}^{3}$Computational Epidemiology
Laboratory, Institute for Scientific Interchange, Torino, Italy}
\affiliation{${}^{4}$Department of Physics, Indiana University, Bloomington, IN 47406, USA}




\begin{abstract}
Among the realistic ingredients to be considered in the computational modeling of infectious diseases, human mobility represents a crucial challenge both on the theoretical side and in view of the limited availability of empirical data. In order to study the interplay between small-scale commuting flows and long-range airline traffic in shaping the spatio-temporal pattern of a global epidemic we {\em i)} analyze mobility data from $29$ countries around the world and find a gravity model able to provide a global description of commuting patterns up to $300$ kms; {\em ii)} integrate in a worldwide structured metapopulation epidemic model a time-scale separation technique for evaluating the force of infection due to multiscale mobility processes in the disease dynamics.  Commuting flows are found, on average, to be one order of magnitude larger than airline flows. However, their introduction into the worldwide model shows that the large scale pattern of the simulated epidemic exhibits only small variations with respect to the baseline case where only airline traffic is considered. The presence of short range mobility increases however the synchronization of subpopulations in close proximity and affects the epidemic behavior at the periphery of the airline transportation infrastructure. The present approach outlines the possibility for the definition of layered computational approaches where different modeling assumptions and granularities can be used consistently in a unifying multi-scale framework.
\end{abstract} 


\maketitle

\section{Introduction}
Computational approaches to the realistic modeling of spatial epidemic spread make use of a wide array of simulation schemes~\cite{riley07-1} ranging from very detailed agent based approaches~\cite{eubank04-1,longini05-1,ferguson05-1,germann06-1,ajelli08-2} to structured metapopulation models based on data-driven mobility schemes at the inter-population level~\cite{rvachev85-1,grais03-2,hufnagel04-1,colizza07-1}. All these approaches integrate a wealth of real world data. However, it is not yet clear how to discriminate the effects of the inclusion/lack of real world features in specific models. This limitation is mainly related to our incomplete knowledge of human interactions and mobility processes, which are fundamental aspects to describe a disease spread. While recent efforts started to make available massive data on human mobility from different sources and at different levels of description~\cite{pentland09, chowell03-1,barrat04-1,guimera05-1,brockmann06-1,viboud06-1,demontis07-1,patuelli07-1,gonzalez08-1}, the multiscale nature of human mobility is yet to be comprehensively explored.
 Human mobility can be generally described by defining a network of interacting communities where the connections and the corresponding intensity represent the flow of people among them~\cite{barrat04-1,wu06-1}. Global mobility networks are made of long range intercontinental air traffic~\cite{barrat04-1,guimera05-1} as well as massive short range commuting flows spanning several orders of magnitude in intensity and spatiotemporal scales~\cite{viboud06-1,demontis07-1,patuelli07-1}. Transportation infrastructures and mobility patterns thus form very complex multiscale networks~\cite{barabasi99-1} which in general elude simple modeling approaches. A multitude of heuristic models for population structure and mobility patterns have been proposed but they all depend on the specific mobility process under consideration~\cite{erlander90-1,ortuzar01-1}.  The limited understanding of the inter-relations among the multiple scales entailed in human mobility and their impact on the definition of epidemic patterns constitute a major road block in the development of predictive large-scale data driven epidemic models. In this context two questions stand out: i) is there a most relevant mobility scale in the definition of the global epidemic pattern? ii) at which level of resolution of the epidemic behavior a given mobility scale starts to be relevant and to which extent? 

To begin addressing these questions, we use high resolution worldwide population data\footnote{The Gridded Population of the World and The Global Rural-Urban Mapping Projects, Socioeconomic Data and Applications Center of Columbia University, http://sedac.ciesin.columbia.edu/gpw} that allow for the definition of subpopulations according to a Voronoi decomposition of the world surface centered on the locations of IATA indexed airports \footnote{International Air Transport Association (IATA), http://www.iata.org}. We have then gathered data on the commuting patterns of $29$ countries in $5$ continents,
constructing short range commuting networks for the defined subpopulations. Extensive analysis of these networks allows us to draw a general gravity law for commuting flows that reproduces commuting patterns worldwide. This law, valid at the scale defined by the tessellation process, is statistically stable across the world due to the globally homogeneous procedure applied to build the subpopulations around transportation hubs. The multiscale networks we obtain are integrated into the GLobal Epidemic and Mobility modeler (GLEaM), a computational platform that uses a metapopulation stochastic model on a global scale to simulate the large scale spreading of influenza like illnesses (ILIs). We develop a time-scale separation technique for evaluating the force of infection due to different mobility couplings that enables us to fully consider the effect of multiscale mobility processes in the disease dynamics. 
The model is then used to simulate a pandemic like event with tunable reproductive ratios. The results obtained from the full multiscale mobility network are compared to the simulations in which only the large scale coupling of the airline transportation network is included. Our analysis shows that while commuting flows are, on average, one order of magnitude larger than the long range airline traffic, the global spatio-temporal patterns of disease spreading are mainly determined by the airline network. Short range commuting interactions  have on the other hand a role in defining a larger degree of synchronization of nearby subpopulations and specific regions which can be considered weakly connected by the airline transportation system. In particular, it is possible to show that short range mobility has an impact in the definition of the subpopulation infection hierarchy. The techniques developed here allow for an initial understanding of the level of data integration required to obtain reliable results in large scale modeling of infectious diseases.

\begin{figure*}[t]
\begin{center}
\includegraphics[width=18cm]{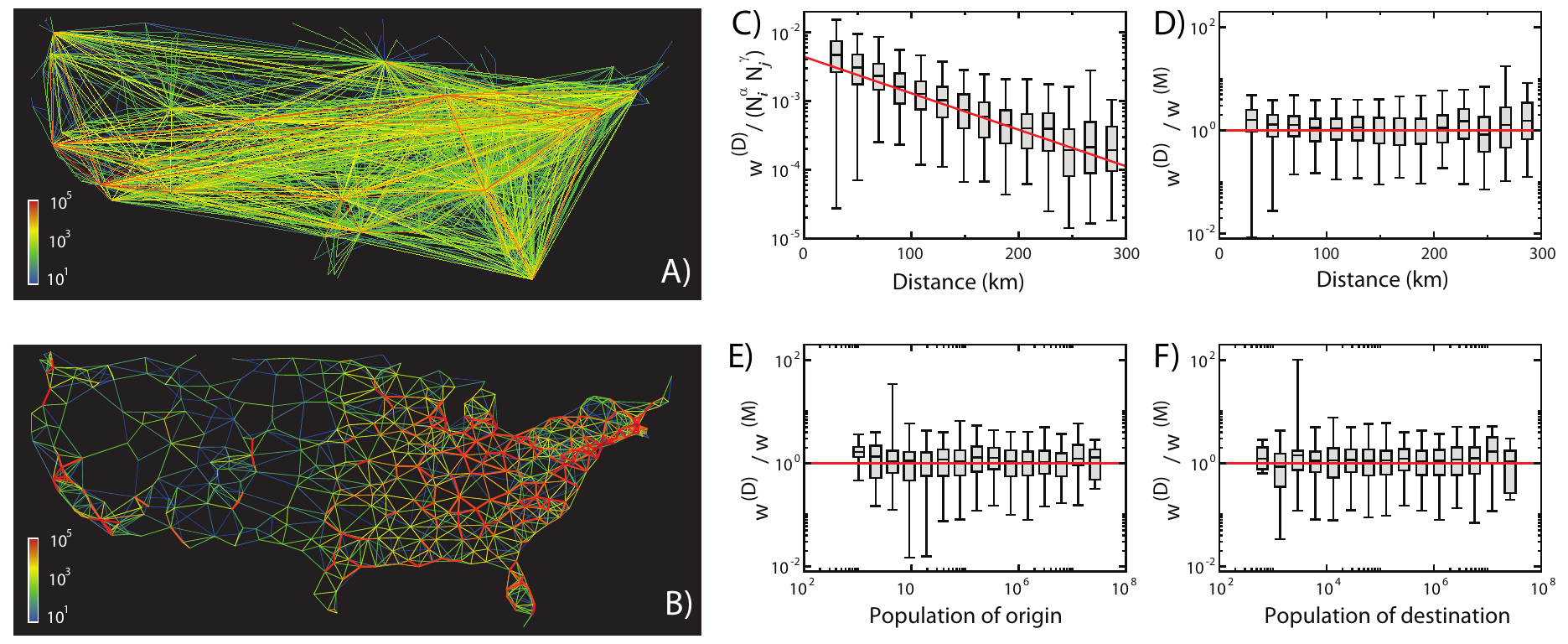}
\end{center}
\caption{\label{Multiscale} Multiscale mobility networks and gravity law fit. A) Continental US airline transportation network. B) Continental US commuting network. The width and color (from blue to red) of the edges represent on a logarithmic scale the intensity of the mobility flow. C) Commuting flux obtained from data ($w^{(D)}$) rescaled by the gravity law's dependence on origin and destination populations ($N_i^{\alpha}N_j^{\gamma}$), as a function of the distance between subpopulations. The number of people commuting between different urban areas decreases exponentially with distance up to $300$ kms. D)-E)-F) Ratio of commuting flux obtained from data ($w^{(D)}$) to corresponding commuting flux predicted by the gravity model with fitted parameters ($w^{(M)}$), as a function of distance, population of origin and population of destination, respectively. The three plots provide values spread around $1$, showing that the synthetic networks generated by the functional form (see Table 1) reproduce well the  commuting fluxes obtained from data. Solid lines in all panels are guides to the eye.}\end{figure*}

\section{Model Description}
Simulations of worldwide epidemic spread are generally based on structured metapopulation models that consider data-driven schemes for long range  mobility  at the inter-population level coupled with coarse-grained techniques within each subpopulation~\cite{rvachev85-1,grais03-2,hufnagel04-1,colizza07-1,flahault91-1,cooper06-1,epstein07-1}.  In this paper, we use the GLobal Epidemic and Mobility  (GLEaM) computational scheme based on a georeferenced metapopulation approach in which the world is partitioned into geographical census regions coupled by population movements. This defines  a subpopulation network where the connections between subpopulations represent the  fluxes of individuals due to the transportation infrastructures and mobility patterns. Inside each subpopulation, the evolution of the infection is described by compartmental schemes in which the discrete stochastic dynamics of the individuals among the different compartments depends on the specific etiology of the disease considered (see Material and Methods). 

\subsection{Multiscale mobility networks} The basic structure of GLEaM includes more than $3,000$ airports in $220$ different countries and the corresponding associated subpopulations. The full consideration of the multiscale nature of mobility networks requires a detailed spatial reference for each subpopulation in such a way as to map short range commuting fluxes among them with sufficient detail. For this reason we use a high resolution population database that allows for the definition of geographical census regions according to a Voronoi decomposition of the world surface centered on the IATA airports. Such subpopulations are not limited by administrative subdivisions or built on postal zone boundaries. The high level of geographical resolution of the  subpopulation database enable us to introduce effective short range mobility due to commuting patterns between neighboring census regions (see Material and Methods). The main difficulty in introducing the effect of commuting flows worldwide is the lack of a global database as opposed to the case of the air traffic flow. Data are scattered in different national and international databases that use different administrative and geographical granularities, and several definitions of commuting flows. We have collected commuting data from $29$ countries in $5$ different continents. Each dataset was mapped into the GLEaM Voronoi tessellation constructing the commuting networks at the subpopulation level. 

In Fig.~\ref{Multiscale}, we show the commuting network of the continental US as obtained by mapping the county commuting data onto the subpopulations used by GLEaM. Commuting data do not consider airline flows that are accounted for by the IATA dataset. On the same scale, we also report the airline traffic network, readily highlighting the difference in scale and spatial structure of the two networks. The commuting network appears as an almost grid-like lattice connecting neighboring subpopulations, while the airline traffic network is dominated by long range connections. The wide range of scales is evident also in the intensities of the mobility flows, spanning several orders of magnitude, with the average commuting flow being one order of magnitude larger than the average airline traffic flow. 
Finally, it should be noted that, in general, commuting flows refer to round trip processes with a characteristic time of the order of $1/3$ days (average duration of a work day) compared with much longer characteristic times for airline travel (average value around $2$ weeks\footnote{Travel Trends 2007, Office for National Statistics, www.statistics.gov.uk}). 

To gain more insight at the global level of the commuting flow we use the general gravity model from transportation theory~\cite{erlander90-1,ortuzar01-1, krings09-1} as a starting point. This assumes that the commuting flow $w_{ij}$ between subpopulation $i$ (with population $N_i$) and subpopulation $j$ (with population $N_j$) takes on the form:
\begin{equation}
w_{ij}= C \frac{N_i^\alpha N_j^\gamma}{f(d_{ij})},
\end{equation}
where $C$ is a proportionality constant, $\alpha$ and $\gamma$ tune the dependence with respect to each subpopulation size and $f(d_{ij})$ is a distance dependent functional form. Gravity laws usually consider power or exponential laws for the behavior of $f(d_{ij})$. The results reported in the literature are variable and generally depend on the way the subpopulations are defined. In our case, we can take advantage of the statistical similarity of the subpopulations centered on major transportation hubs. We tested the gravity law by the statistical analysis of more than $10^4$ 
flows worldwide and we found that the best fit is obtained by using an exponential function $f(d_{ij})= \mathrm{exp}(d_{ij}/r)$ where $r$ is the 
characteristic length that governs the decay of commuting flows. In Table 1, we report the estimated values for the exponents 
$\alpha$, $\gamma$ and $r$. 

\begin{table}
\begin{center}
\begin{tabular}{lccc} 
\hline\hline
$d\;(km)$ & $\alpha$ & $\gamma$ & $r \, (km)$\\
\hline\hline
$\le 300$ & $0.46 \pm 0.01$ &    $0.64 \pm 0.01$ & $82 \pm 2$ \\
$ > 300$  & $0.35 \pm 0.06$ & $0.37\pm0.06$    & NA \\
\hline
\hline
\end{tabular}
\caption{Exponents of the gravity law as obtained by applying a multivariate analysis to global commuting data. These values are used in the construction of synthetic commuting network worldwide.}
\end{center}
\end{table}

Noticeably, we can validate the gravity law at the level of each country as shown in Fig.~\ref{Multiscale}, where the commuting flows obtained from data are compared to the synthetic ones predicted by the model, as  functions of the various variables of the gravity law. It is worth remarking that while the present gravity law works well at the granularity defined by our Voronoi decomposition, these results cannot be extrapolated at different granularities. For instance, we notice that the exponents obtained by our approach are quite close to those obtained by Viboud et al.~\cite{viboud06-1}, although the spatial decay has a completely different form. It is difficult to ascertain the origin of the difference, but it must be noted that we are working with subpopulations which are by construction statistically more homogeneous and of larger size than the county level used in Ref.~\cite{viboud06-1}. Administrative regions might indeed impose  boundaries that define subpopulations not clearly associated to centers of gravity for mobility processes, as e.g. large urban areas cut in multiple counties.

The above gravity law allows us to work with two different worldwide commuting networks. An entirely synthetic one, generated using the gravity law fitted to the empirical data, and one integrating the empirical data. In the following, we will report the results obtained only with the synthetic network as there are no significant differences are observed when we compare the results obtained by using the synthetic and the real data networks.

\subsection{Epidemic simulations} To study the effect that commuting networks have on the overall spread of an emerging disease we consider the simulation of an ILI  starting in Hanoi  and compare the results with a simulation in which we include only airline traffic as in previous works~\cite{colizza07-1}. The model includes a seasonal dependence of the transmission and simulations are assumed to start on April 1.

GLEaM is fully stochastic and takes into account the discrete nature of individuals both in the travel coupling and in the compartmental transitions. The transmission model within each urban area follows a compartmentalization specific to the disease under study. Here we use the classic ILI compartmentalization in which each individual is classified by one of the discrete states such as susceptible ($S$), latent ($L$), infectious ($I$) and permanently recovered ($R$). Infectious persons are further subdivided into asymptomatic, symptomatic traveling and symptomatic non-traveling as detailed in the Material and Methods. The discrete nature of individuals is implemented by introducing binomial and multinomial processes inside each urban area for the stochastic evolution of the infection. 

In GLEaM, the airline mobility is integrated explicitly as individuals are allowed to travel from one subpopulation to another by means of the airline transportation network~\cite{colizza06-1} similarly to the models in Refs.~\cite{rvachev85-1,grais03-2} and the stochastic generalizations of Ref.~\cite{hufnagel04-1}. In each subpopulation $j$ the number of individuals is $N_j(t)$ and $X_j^{[m]}(t)$ is the number of individuals in compartment $[m]$ of the disease evolution at time $t$, therefore $N_j(t) = \sum_{m} X_j^{[m]}(t)$.  The dynamics of individuals traveling between cities is described by the stochastic transport operator $\Omega_j (\{X^{[m]}\})$ representing the net balance of individuals in a given state $[m]$ that entered or left each city $j$. This operator is a function of the traffic flows per unit time $\omega_{j\ell}$ with the neighboring cities   and of the city population $N_j$. In particular, the number of passengers of each category traveling from city $j$ to city $\ell$ is an integer random variable, in that each of the potential travelers has a probability $p_{j\ell}=\omega_{j\ell}\, \delta t / N_j$ to go from $j$ to $\ell$ in the time interval $\delta t$. In city $j$, the number of passengers traveling on each connection $(j,\ell)$ at time $t$ defines a set of stochastic variables which follow a multinomial distribution~\cite{colizza07-1}. The calculation can be extended to include transit traffic up to one connection flight~\cite{colizza06-2}.

The introduction of commuting flows represents a numerical challenge as it acts at a very different time scale with individuals which have very short visit duration in the neighboring subpopulation. While the airline traffic finds a natural time scale of one day, an equivalent mechanistic simulation of the commuting mobility would require to work on a much smaller time scale hardly compatible with the airline traffic. This problem can be solved by relying on the results of~\cite{keeling02-1}. Commuting flows govern subpopulation interactions by defining the visiting rate of individual in subpopulation $i$ to subpopulation $j$ as $\sigma_{ij}=w_{ij}/N_i$. Visiting individuals have a very short visit time associated to high return rates, $\tau$, to their original subpopulations ($\tau \ge 3 \; day^{-1}$), corresponding to  $\sigma_{ij}  \ll \tau$ for all the populations. It can be shown that the system is then described by stationary quantities $X_{ij}^{[m]}$ in each compartment $[m]$ for time scales larger than $\tau^{-1}$ (the time scale governing the relaxation time to equilibrium across subpopulations) in which $X_{ij}^{[m]}$ is defined as the number of visitors currently in $j$ coming from subpopulation $i$ in compartment $[m]$. When the disease duration is significantly larger than $\tau^{-1}$, as it is the case for the ILI with characteristic time $\mu^{-1}\simeq 3 \; day$, we can use a time scale separation approximation in which at each time step the force of infection experienced by susceptible individuals in each subpopulation is a function of the stationary values $I_{ii}$ and $I_{ji}$. The full force of infection used in the model is reported below.

\begin{figure}[t]
\vspace{-0.5cm}
\begin{center}
\includegraphics[width=18cm]{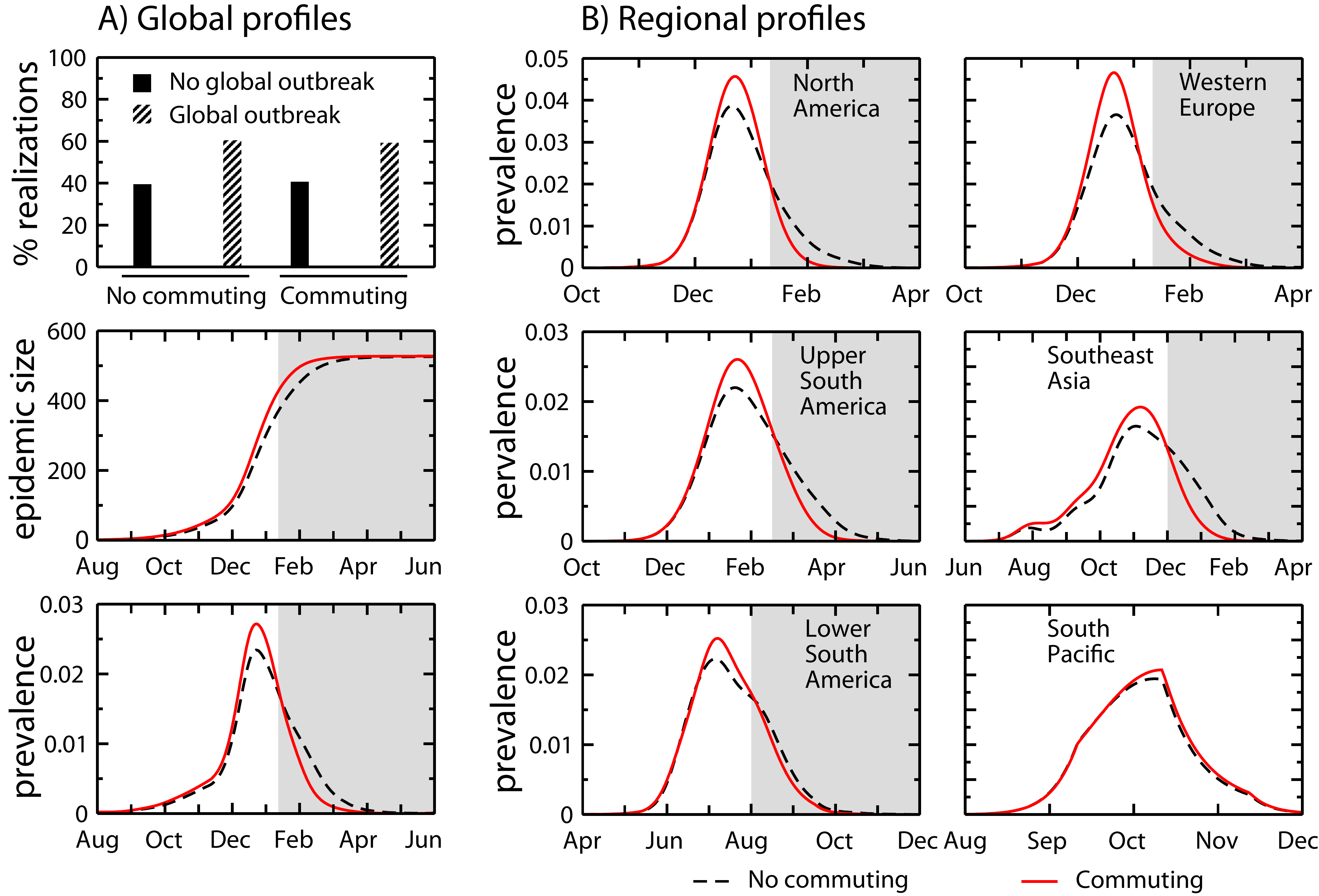}
\end{center}
\caption{\label{Global_impact} Comparison of GLEaM predictions at the global and regional level obtained with and without commuting. Results refer to  a pandemic influenza with $R_0=1.9$ starting in Hanoi on April $1st$. A) Top: Probability of outbreak. About $40\%$ of the realizations leads to an extinction at the source (Hanoi), whereas the remaining $60\%$ causes a pandemic reaching more than $100$ countries (i.e. a global outbreak). Center-Bottom: global profiles for the epidemic size and the prevalence, averaged over global outbreaks. B) Regional profiles  for the epidemic size and the prevalence, averaged over all runs that led to an outbreak in the given region. All results show a very limited impact of the commuting on the simulated patterns, more evident in the faster decay in the prevalence profiles as highlighted by the shaded areas. Reported results are averaged over $10^3$ outbreak realizations.}
\end{figure}
Fig.~\ref{Global_impact} shows the global and regional impact of a pandemic influenza with $R_0=1.9$ starting in Hanoi on April 1 and compare the epidemic profiles with and without commuting flows. The results show that the overall timing and size of the epidemic is weakly affected by also considering the commuting network in GLEaM.  Both the probability of having a global outbreak and the overall profiles are very similar in the two cases, with the global epidemic size  at the end of the first year almost unaffected by the inclusion of commuting. We perform a sensitivity analysis by testing the outset variation of the intensity of the commuting flows and varying the return time rate $\tau$ of more than one order of magnitude. In the Supporting Information we show that the profiles do not show significant variations, the results being very robust against strong fluctuations in the commuting mobility process.

\begin{figure}[t]
\vspace{-1.cm}
\begin{center}
\includegraphics[width=18cm]{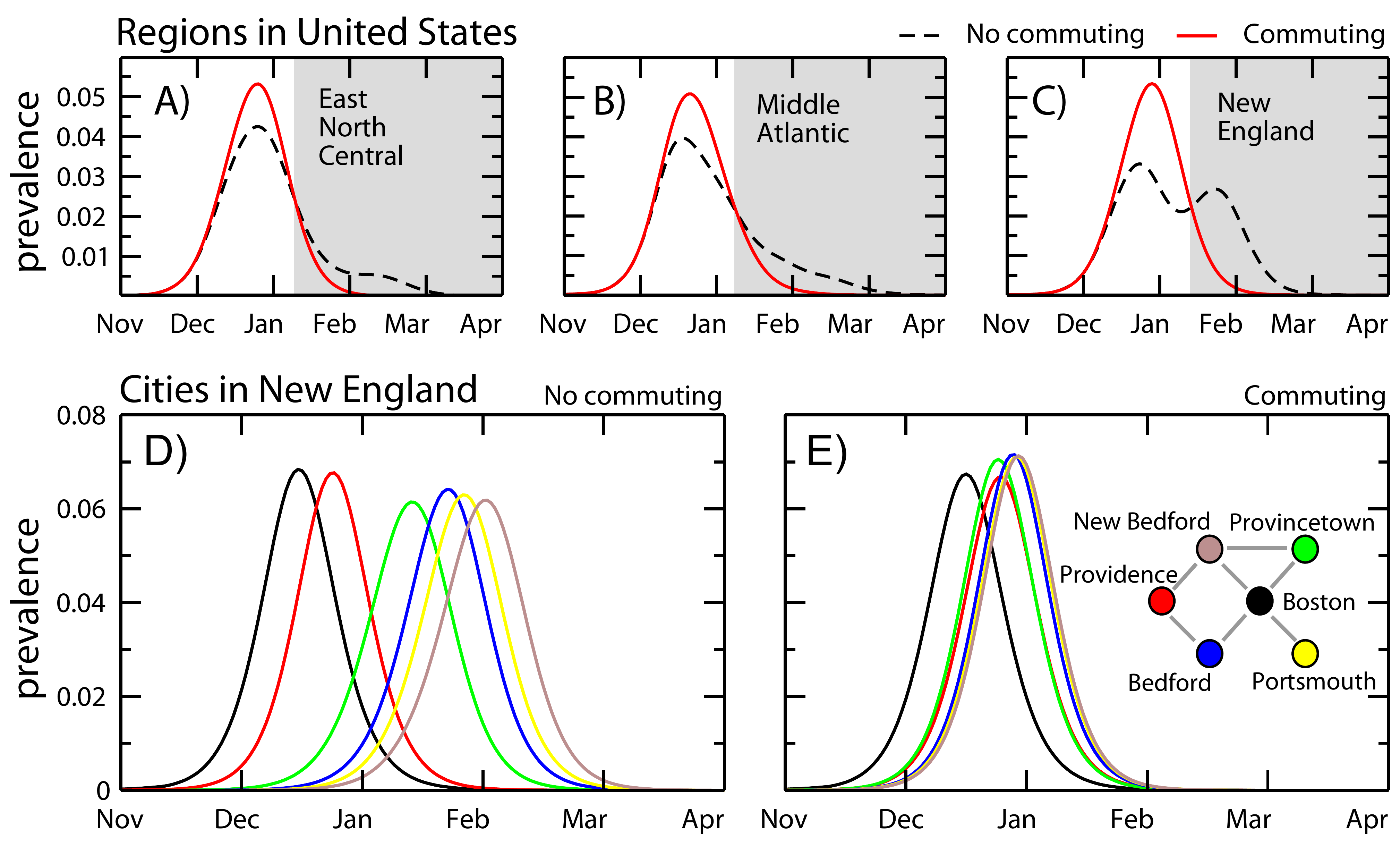}
\end{center}
\caption{\label{Sync} Comparison of GLEaM predictions at the local level obtained with and without commuting. A)-B)-C) Prevalence profiles of $3$ continental US regions. The effect of commuting is visible in the faster decay (as highlighted by the shaded areas) and absence of multiple peaks. D)-E) Prevalence profiles for Boston area and the surrounding cities with no commuting (D) and with commuting (E). A schematic network representation of the short-range connections is shown for guidance. The synchronization among the prevalence profiles is considerably increased when commuting is considered, with a reduction of over one month in the time interval between  peaks in  neighboring cities. Reported profiles are averaged over $10^3$ outbreak realizations.}
\end{figure}

The effect of commuting flows is, however, noticeable during the tail of the epidemic event. 
As presented in Fig.~\ref{Global_impact}, many regions of the world show a broader tail in the absence of commuting, showing that the commuting coupling enhances the synchronization of the local epidemic profiles. The observed broadening of an epidemic profile that includes multiple subpopulations is due to the different timing of the outbreak that reaches the various subpopulations. The effect  is more pronounced in the lack of short range coupling, as  highlighted in the example reported in Fig.~\ref{Sync}D-E of an air transportation hub loosely connected by air travel flow to the surrounding subpopulations. As expected, no significant change is observed in the hub profile, whereas the time delay in neighboring locations with limited airline connections is dramatically reduced by the coupling due to local commuting flows. After infecting the hub, the epidemic radiates out to the neighboring geographical census areas in a pattern reminiscent of the physical process of diffusion. This effect naturally leads to a much stronger correlation and synchrony in the evolution of the pandemic at the local level.

\begin{figure*}[t]
\begin{center}
\includegraphics[width=18cm]{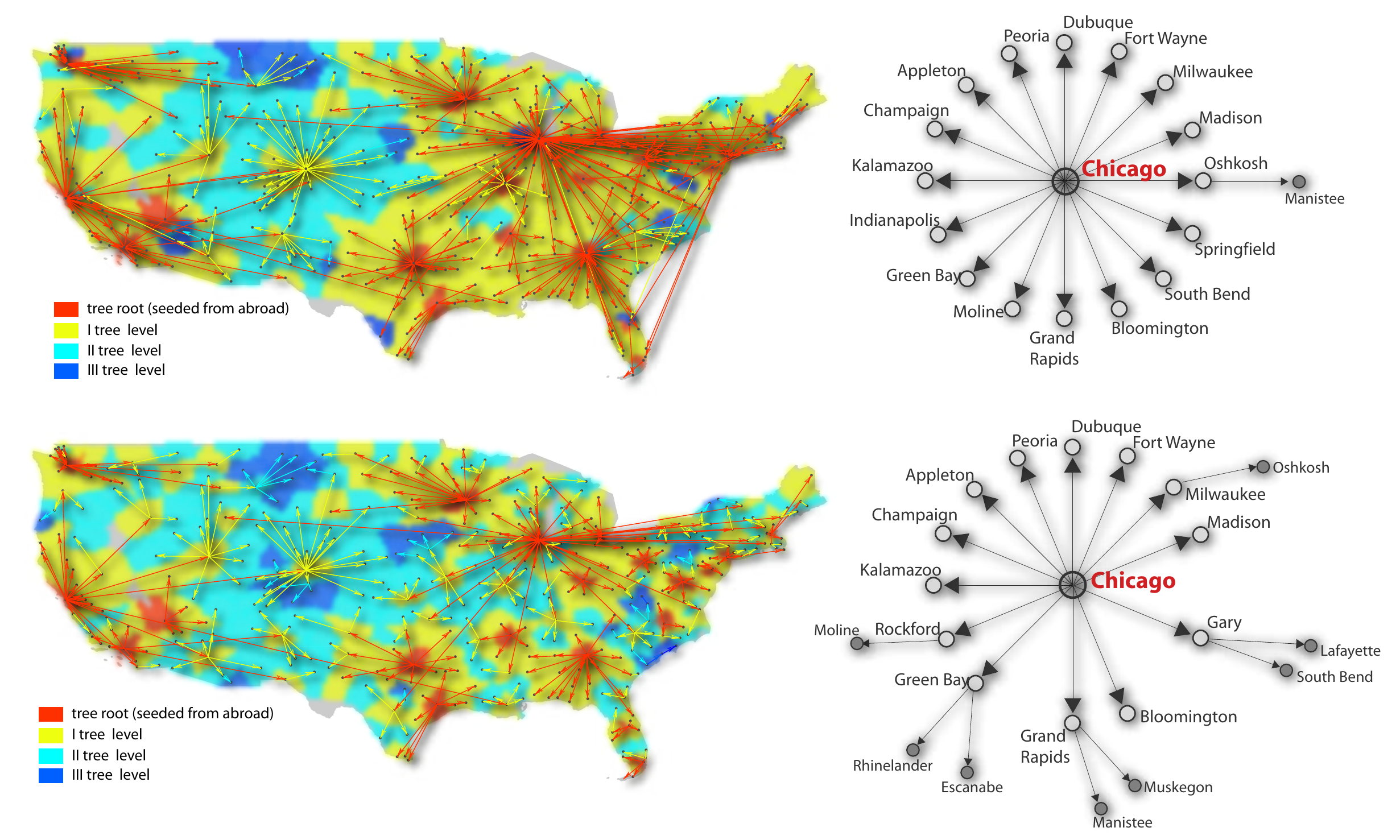}
\end{center}
\caption{\label{causality} Geographical representation of the continental US epidemic invasion tree A) with only airline traffic and B) when both airline traffic and commuting are considered. Red represents the roots, i.e. the first cities that were seeded from abroad, and as we move down the tree the colors change from yellow to dark blue. The arrows representing the edges of the tree are colored as the parent node. We also provide a schematic representation of the invasion tree rooted at Chicago when only flights are considered C) and with both air traffic and commuting D). As demonstrated in both examples, the spreading pathway is completely dominated by the airline hubs as the only sources of imported seeds. However, the hierarchy is broken by the introduction of commuting flows as the number of shells around the airline hubs and the branches at the secondary nodes increase.}
\end{figure*}

Commuting flows therefore alter the hierarchy of epidemic transmission from region to region. This hierarchical organization can be inferred by constructing the epidemic invasion tree that represent the  transmission of the infection from one subpopulation to the other during the history of the epidemic. The stochastic nature of the epidemic process implies that each realization will produce a different tree. An overall epidemic invasion network can be constructed by defining weighted directed links, $T_{ij}$, that denote the probability that the epidemic in subpopulation $j$ is seeded by individuals belonging to the subpopulation $i$. This probability is defined by the ratio between the number of realizations in which we have a seeding $i\to j$  and the total number of realizations. When constructing the epidemic invasion tree we use averages over $10^3$ realizations. Finally, in order to highlight only the most likely infection tree we construct the minimum spanning  tree from the world seeding subpopulation where we minimize the distance defined on each link as $T^{-1}_{ij}$. In Fig.~\ref{causality}, we show the infection arrival time hierarchy in the two considered scenarios for the continental United States.  In the absence of commuting (panel A), airline hubs have a predominant role and are completely responsible for spreading the disease to every other location through direct air connections. This leads to the counterintuitive effect that locations near a large airport, but with no frequent direct flight to that airport, can be infected only much later through a convoluted sequence of flights. On the other hand, when we superimpose the commuting network we obtain the expected effect of reducing the importance of large airports and increasing the locality of the epidemic spread (see panel B). The inclusion of commuting patterns is therefore relevant in the evaluation of the epidemic invasion path and timing. 

\section{Voronoi tessellation around main transportation hubs} 
Here we use the population database of the "Gridded Population of the World" project of SEDAC (Columbia University), that estimates the population with a granularity given by a lattice of cells covering the whole planet at a resolution of $15 \times 15$ minutes of arc. We define the geographical census areas centered around IATA airports by assigning each cell to the closest airport within the same country as long as the distance between the center of the cell and the airport is lower than $200 \,kms$. This is the characteristic length scale of the cell-airport distribution as well as a tipping point for the intensity of the ground commuting flows. Such a procedure defines a Voronoi-like tessellation for the populated cells of the world (see the Supplementary Information for further details). 

\section{Disease structure, seasonality and $R_0$}
In each urban area, the evolution of the disease is governed by the compartmental scheme of the baseline scenario of Ref.~\cite{colizza07-1}. A susceptible  individual $\left(S\right)$ in contact with a symptomatic  ($I^{t},I^{nt}$, traveling or not-traveling, respectively)  and asymptomatic $\left(I^a\right)$ infectious individual contracts the infection at rate $\beta$ or $r_\beta \beta$, respectively, and enters the latent $(L)$ compartment, where he/she is infected but not yet infectious. At the end of the latency period, individuals in the latent class enter one of the symptomatic infectious compartments ($I^{t},\,I^{nt}$) with probability $1-p_a$ or become asymptomatic $\left(I^a\right)$ with probability $p_a$. Symptomatic individuals are further divided between those who are allowed to travel $\left(I^t\right)$ with probability $p_t$ and those who are prevented from doing so $\left(I^{nt}\right)$ with probability $1-p_t$, dependending on the severity of symptoms. All infectious individuals enter the permanently recovered/removed compartment $(R)$ at a rate of $\mu$ per day. The latent period has an average duration of $\varepsilon^{-1}=1.9$ day and is assumed to be followed by an infectious period with a mean duration of $\mu^{-1}=3$ day~\cite{colizza07-1,longini04-1,longini05-1}. Given that infection has occurred, we assume that individuals become asymptomatic with probability $p_a=0.33$~\cite{colizza07-1,longini04-1,longini05-1}. The relative infectiousness of asymptomatic individuals is $r_\beta=0.5$~\cite{colizza07-1} and symptomatic individuals are allowed to travel with probability $p_t=0.5$. 
The contagion process (i.e. the generation of new infectious through the transmission of the disease from infectious to susceptible individuals) and the spontaneous transitions (e.g. from latent to infectious or from infectious to recovered) are modeled with binomial and multinomial distributions (see the Supplementary Information for a detailed descritpion of the processes). 
The threshold parameter of the disease that determines the spreading rate of infection is called basic reproduction number $\left(R_0\right)$, and is defined as the average number of infected cases generated by a typical infectious individual when introduced into a fully susceptible population~\cite{may92-1}. For our compartmental model we have  $R_0=\beta \mu^{-1} [1-p_a + r_\beta p_a]$.
The $R_0$ values indicated in the figures and discussed in the paper do not consider the  effect of seasonality and the commuting in the force of infection.
We take into account the seasonal behavior of influenza, by adopting the scheme from Ref.~\cite{colizza07-1}. The transmission rate $\beta_j$ in each geographical census area is adjusted by a scaling factor which varies monthly according to the city's climatic zone. For example, cities in the tropical zone have a scaling factor that is always $1$ independent of the season. See Ref.~\cite{colizza07-1} and its supporting information for details.

\section{Commuting short range couplings} 
The effect of commuting in the spread of infection can be considered implicitly by evaluating the force of infection between subpopulations coupled by commuting flows. In particular, the number of individuals $X_j^{[m]}(t)$ in each compartment $[m]$ at time $t$ in city $j$ can be expressed as the sum of individuals $X_{jj}^{[m]}(t)$ who are actually present in their home subpopulation and those $X_{ji}^{[m]}(t)$ who are visiting a neighboring city $i$~\cite{sattenspiel95-1}. By definition it follows that $X_{j}^{[m]}(t) = X_{jj}^{[m]}(t) + \sum_{ i \in \upsilon(j) } X_{ji}^{[m]}(t)$, where $\upsilon(j)$ denotes the set of neighbors of $j$. All individuals in each traveling compartment  visit a neighboring subpopulation at a rate of $\sigma_{ij}$ for an average duration of $\tau^{-1}$. It is possible to show that at the equilibrium the populations $X_{jj}^{[m]}$ and $X_{ji}^{[m]}$ can be expressed as:
\begin{equation}
X_{jj}^{[m]} = {X_j^{[m]} \over (1+\sigma_j/\tau)}  \; {\rm and } \; X_{ji}^{[m]} = {X_j^{[m]} \over (1+\sigma_j/\tau)} \sigma_{ji} / \tau \;, \label{pop}
\end{equation}
where $\sigma_j=\sum_{ i \in  \upsilon(j) } \sigma_{ji}$ denotes the total commuting rate of $j$. Whereas $X_{jj}^{[m]}=X_j^{[m]}$ and $X_{ji}^{[m]}=0$ for all the other compartments which are restricted from traveling. In the case of $\tau \gg \sigma_i$, the relaxation time to equilibrium values is dominated by the return rate $\tau$ to home, as it is the case of commuting. We can use the equilibrium values of population sizes in our calculations of force of infection since the $\tau^{-1}$ is much smaller than the time scales of disease evolution (i.e. $\varepsilon^{-1}$ and $\mu^{-1}$). 
New infections in a subpopulation are due to the transmission between susceptibles and infectious occuring in the subpopulations or during a visit to a neighboring subpopulation. Taking this into account, it is possible to derive the force of infection $\lambda_j$ in $j$ as:
\begin{eqnarray}
\lambda_ j &=&  {\beta_j \over (1+\sigma_j/\tau) N_j^\ast} \left[ I_j^{nt} + {I_{j}^{t} + r_\beta I_{j}^a \over  1+\sigma_j/\tau} \right] \nonumber \\ 
&+&  {1 \over (1+\sigma_j/\tau) \tau}  \sum_{ i \in \upsilon(j) } \left[ {\beta_j \sigma_{ij} \over N_j^\ast }  {I_{i}^{t} + r_\beta I_{i}^a \over 1+\sigma_i/\tau} \right. \\
&+&  \left. {\beta_i \sigma_{ji} \over N_i^\ast} \left( I_i^{nt} + {I_{i}^{t} + r_\beta I_{i}^a \over 1+\sigma_i/\tau}  +  \sum_{ \ell \in \upsilon(i) } { \sigma_{\ell i}  \over \tau} { I_{\ell}^{t} + r_\beta I_{\ell}^a \over 1+\sigma_\ell/\tau } \right) \right]  \nonumber \;.
\end{eqnarray} \label{finf}
A detailed derivation is provided in the Supporting Information.
In the above expression $N_{j}^{\ast}=N_{jj} + \sum_{i \in \upsilon(j)} N_{ij}$ is the actual number of individuals in $j$ due to commuting. The first terms of the r.h.s. of Eq.~(\ref{finf}) takes into account the transmission of the infection from the local infectious individuals in $j$. The second term considers the transmission due to the infectious individuals during their visits to $j$ with local susceptible persons. The third and forth terms consider the interactions of susceptible individuals during their visits to neighboring subpopulations $i$ with the local infectious persons and the infectious visitors of $i$, respectively. Here we have also considered that the transmission rate $\beta$ may be different in each population. The last expression includes second order commuting terms (e.g. $\sigma_{ji} \sigma_{\ell i} / \tau^2$), which are neglected in the actual computation. The probability of new infections to be generated in city $j$ is finally given by $\lambda_j \delta t$ in the time interval $\delta t$, acting on a pool of susceptible individuals $S_j$. 

\section{Conclusions}

Data collected from $29$ countries in $5$ continents were used to fit a gravity law that was then used to model commuting behavior between the Voronoi geographical census areas built around every airport indexed by IATA. The effect of adding this short range commuting network to a worldwide epidemic model including all airline traffic flowing among $3362$ airport locations allows us to discriminate the main contribution of the long and short-range mobility flows. The impact of the epidemic does not change as the competition between the long and short-range coupling acts only at the beginning of the epidemic in each subpopulation. Both coupling terms become a second order effect once the epidemic ramps up and the major force of infection is endogenous to the subpopulation. Therefore,  both coupling mechanisms affect just the  hierarchy of epidemic progression and its timing. 
On the one hand, the global epidemic behavior is governed by the long-range airline traffic that determines the arrival of infectious individuals on a worldwide scale. At the local level, however, the short range epidemic coupling induced by commuting flows creates a synchrony between neighboring regions and a local diffusive pattern with the epidemic flowing from subpopulations with major hubs to their geographical proximity.
These results clearly show that the level of detail on the mobility networks can be chosen according to the scale of interest. Neglecting local coupling for instance does not produce a dramatic effect if one is mainly interested in the global overall pattern at the granularity level of a large geographical area or country. On the other hand, more refined strategies that require access to finer granularity can be implemented by the progressive addition of details without radically altering the perspective achieved at the larger scales. This is extremely important in the balance between computational time and flexibility of models, and becomes very relevant when computational approaches are used in  real time  to aid the decision process for a public health emergency, since  new data may need to be incorporated in real time. The present analysis opens the path to quantitative approximation schemes which calibrate the level of data resolution and the needed computational resources with respect to the accuracy in the description of the epidemics. In the approach we present, we move a first step in this direction by defining the time-scale separation method for local couplings. 

\section*{Acknowledgments}
We are grateful to the International Air Transport Association for making the airline commercial flight database available to us. This work has been partially funded by the NIH R21-DA024259 award, the Lilly Endowment grant 2008 1639-000 and the DTRA-1-0910039 award to AV; the EC-ICT contract no. 231807 (EPIWORK) to AV and VC; the ERC Ideas contract n.ERC-2007-Stg204863 (EPIFOR) to VC.



\bibliographystyle{pnas}


\end{document}